\documentclass[fleqn,usenatbib]{mnras}

\usepackage{newtxtext,newtxmath}

\usepackage[T1]{fontenc}

\DeclareRobustCommand{\VAN}[3]{#2}
\let\VANthebibliography\thebibliography
\def\thebibliography{\DeclareRobustCommand{\VAN}[3]{##3}\VANthebibliography}

\usepackage{graphicx}	
\usepackage{amsmath}	

\newcommand{\rc}{\color{black}}
\newcommand{\rcc}{\color{black}}

\title[Photoevaporation vs. Core-powered Mass-loss]{Mapping out the parameter space for photoevaporation and core-powered mass-loss}

\author[Owen, J. E. \& Schlichting, H. E.]{
James E. Owen$^{1,2}$\thanks{E-mail: james.owen@imperial.ac.uk} and Hilke E. Schlichting$^{2}$
\\
$^{1}$ Astrophysics Group, Department of Physics, Imperial College London, Prince Consort Rd, London SW7 2AZ, UK\\
${2}$ Department of Earth, Planetary, and Space Sciences, University of California, Los Angeles, CA 90095, USA
}

\date{Accepted XXX. Received YYY; in original form ZZZ}

\pubyear{2015}

\begin{document}
\label{firstpage}
\pagerange{\pageref{firstpage}--\pageref{lastpage}}
\maketitle

\begin{abstract}
Understanding atmospheric escape in close-in exoplanets is critical to interpreting their evolution. We map out the parameter space over which photoevaporation and core-powered mass loss dominate atmospheric escape. Generally, the transition between the two regimes is determined by the location of the Bondi radius (i.e. the sonic point of core-powered outflow) relative to the penetration depth of XUV photons. Photoevaporation dominates the loss when the XUV penetration depth lies inside the Bondi radius ($R_{XUV}<R_B$) and core-powered mass-loss when XUV radiation is absorbed higher up in the flow ($R_B<R_{XUV}$). The transition between the two regimes occurs at a roughly constant ratio of the planet's radius to its Bondi radius, with the exact value depending logarithmically on planetary and stellar properties. In general, core-powered mass-loss dominates for lower-gravity planets with higher equilibrium temperatures, and photoevaporation dominates for higher-gravity planets with lower equilibrium temperatures. However, planets can transition between these two mass-loss regimes during their evolution, and core-powered mass loss can ``enhance'' photo-evaporation over a significant region of parameter space. Interestingly, a planet that is ultimately stripped by core-powered mass-loss has likely only ever experienced core-powered mass-loss. In contrast a planet that is ultimately stripped by photoevaporation could have experienced an early phase of core-powered mass-loss. Applying our results to the observed super-Earth population suggests that it contains significant fractions of planets where each mechanism controlled the final removal of the H/He envelope, although photoevaporation appears to be responsible for the final carving of the exoplanet radius-valley.
\end{abstract}




\section{Introduction}
The discovery of the first close-in exoplanet around a main-sequence star (the hot Jupiter 51 Peg b, \citealt{Mayor1995}) led to speculation that atmospheric escape may be important for driving the evolution of a planet's bulk properties \citep{Burrows1995,Baraffe2004}. While it is now firmly established that atmospheric escape alone cannot cause sufficient mass-loss to affect the bulk of hot Jupiters \citep[e.g.][]{Hubbard2007}, the discovery of lower mass planets gave rise to the idea that their primordial hydrogen dominated atmosphere could lose a significant fraction, or all their mass over a planet's billion year lifetime \citep[e.g.][]{Valencia2010,Owen2012,Lopez2012,Ginzburg2018}. 

Under the assumption that all close-in, low-mass planets were born with voluminous hydrogen-dominated atmospheres that they accreted from their parent nebula, \citet{Owen2013} studied the impact of atmospheric escape at the population level. This work demonstrated that atmospheric escape carved distinct features in the exoplanet population: firstly, the ``evaporation desert'' - a lack of intermediate (2-6~R$_\oplus$) sized planets close to their host star, matching the observed hot Neptune desert \citep[e.g.][]{Szabo2011,Lundvist2016,Mazeh2016}; and secondly a ``evaporation valley'' where planets which retained approximately $\sim 1\%$ hydrogen by mass are separated in radius-period (and density) space from those planets that completely lost their atmospheres ending up as ``stripped cores''. This evaporation valley bears remarkable similarity to the observed exoplanet radius-gap initially identified in Kepler data \citep[e.g.][]{Fulton2017,VanEylen2018,Fulton2018,Ho2023}. These two features are generic in a planet population born with a hydrogen-dominated atmosphere undergoing atmospheric escape \citep[e.g.][]{Lopez2013,Jin2014,Chen2016,Ginzburg2018,Gupta2019,Wyatt2020}; and are well understood to be a consequence of (any) efficient atmospheric escape and the mass-radius relationship for these planets \citep[e.g.][]{Owen2017,Gupta2019,Mordasini2020}. 

However, the details of the atmospheric escape process do matter. And although different escape models infer broadly similar properties about the underlying exoplanet population, they do differ, for example, in their inferred underlying exoplanet mass-distribution and inferred initial atmospheric mass fractions \citep[e.g.][]{Gupta2019,Gupta2020,Rogers2021,Rogers2021b}. Furthermore, as we progress into the era of exoplanetary characterisation, understanding the details of the escape processes will become paramount since the fractionation of heavy species in a hydrogen-dominated outflow can be extremely sensitive to the details of the escape process \citep[e.g.][]{Zahnle1986}. Thus, while the features imprinted in the exoplanet population due to different escape mechanisms may be subtle, the composition differences of any remaining primordial or secondary atmosphere are likely to be vastly more sensitive to the underlying physics of the escape mechanism \citep[e.g.][]{Misener2021}. 

Currently, atmospheric escape models for close-in exoplanets commonly fall into two generic classes: ``photoevaporation'' where the outflow is driven by heating from X-ray and UV stellar photons \citep[e.g.][]{Lammer2003,Yelle2004,GarciaMunoz2007,MurrayClay2009,Owen2012} and "core-powered mass-loss", where the outflow is driven by heating from the planet's cooling luminosity and stellar bolometric luminosity \citep[e.g.][]{Ginzburg2016,Gupta2019,Gupta2020,Gupta2022}. All the previously discussed evolutionary models only focus on the impact of one of these classes of escape models. The underlying principle of both models remains the same: heating drives a transonic, hydrodynamic outflow akin to a Parker wind \citep[e.g.][]{Parker1958}. However, the outflows' mass-loss rates, temperature and ionization structures can be different, in some cases differing by orders of magnitude. 

These different escape classes are not mutually exclusive: in the absence of XUV irradiation, a {\rc highly irradiated} planet will naturally launch a core-powered outflow; on the other hand, under extreme XUV irradiation, photoevaporation will always occur as demonstrated in other areas of astrophysics \citep[e.g.][]{Begelman1983,Bertoldi1990,Owen2012disc}. Therefore, the more pertinent question is under what conditions does atmospheric escape occur in a photoevaporative manner, and when does core-powered mass-loss occur? There has been some simulation work looking into the transition. When a temperature floor equivalent to bolometric stellar heating to the equilibrium temperature has been included \citep[e.g.][]{Kubyshkina2018a}, a transition between a bolometric outflow (at the equilibrium temperature) and photoevaporation did occur. \citet{Kubyshkina2018a} found that this transition happened for lower values of the ``escape parameter'' (the ratio of a gas particle's binding energy to its thermal energy at the equilibrium temperature), with bolometric outflows were more prevalent when the escape parameter was roughly smaller than 10.  In addition, using the \textsc{aiolos} code that includes explicit XUV and bolometric heating from the star and planetary core, \citet{Schulik2023_new} showed a calculation for a GJ 436 b-like planet that transitioned from core-powered mass-loss to photoevaporation as the XUV flux was increased relative to the bolometric flux. However, the governing physics underlying the transition between photoevaporation and core-powered mass-loss has not been studied, nor has an idea of when and where different escape mechanisms dominate both during an individual planet's lifetime and across the exoplanet population. Thus, in order to guide future expensive radiation-hydrodynamic simulations, here we use (semi-)analytical techniques to lay the physical foundations governing the transition between photoevaporation and core-powered mass-loss.

\section{Problem Construction}
\begin{figure}
    \centering
    \includegraphics[width=\columnwidth]{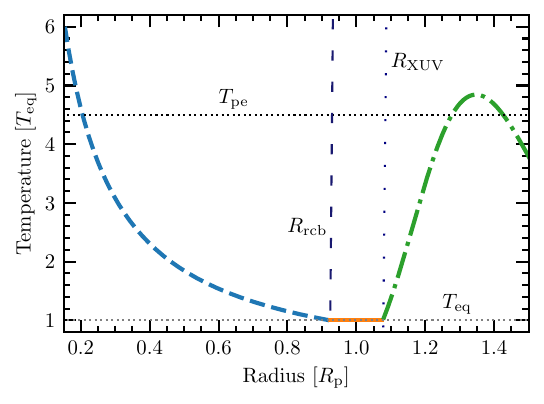}
    \caption{A schematic of the temperature structure of a planet's envelope/atmosphere. This temperature structure is shown in units of the planet's equilibrium temperature.  The temperature structure is adiabatic (thick dashed line) deep in the planet's envelope. At the radiative-convective boundary $R_{\rm rcb}$, it becomes radiative and approximately isothermal. Once the XUV photons can penetrate the atmosphere at $R_{\rm XUV}$, a photoevaporative outflow is launched. We also show $T_{\rm pe}$ a representative XUV heated photoevaporative outflow temperature. The planet's radius ($R_p$) and the optical transit radius lie in between $R_{\rm rcb}$ and $R_{\rm XUV}$. }
    \label{fig:T_schematic}
\end{figure}

In order to gain insights into the problem, we consider the basic structure of a hydrogen-dominated envelope. Figure~\ref{fig:T_schematic} shows a schematic of the temperature structure we are investigating. Deep in the planet's envelope, we model it to be convective, and due to the low internal luminosities, we approximate its structure as adiabatic. The atmosphere becomes radiative at the radiative-convective boundary ($R_{\rm rcb}$). Again, due to the low internal luminosities, we approximate this radiative layer to be isothermal with a temperature set to the planet's equilibrium temperature ($T_{\rm eq}$). It is this isothermal layer that represents the outflowing core-powered mass-loss region. Below the planet's radius ($R_p$, which we define as the $\tau=1$ surface to outgoing thermal -- IR -- radiation), the outflow is mainly powered by the planet's internal luminosity. Above $R_p$, the star's bolometric luminosity provides an additional energy source. XUV photons can penetrate the atmosphere to $R_{\rm XUV}$, which we take to be the $\tau=1$ surface to XUV photons. They heat the rarefied gas to high temperatures and drive a photoevaporative outflow. While only in the particular case of an XUV heated region in recombination-ionization equilibrium is the XUV region typically exactly isothermal, we can gain a lot of insights in the ensuing sections by considering the XUV heated region to be isothermal with a representative temperature $T_{\rm pe}$. {\rc In section~\ref{sec:numerical}, we calculate this representative temperature. However, in this preceding section, we keep it arbitrary but assume it's larger than the equilibrium temperature, guided by the fact simulations indicate it's in the range $3,000-10,000$~K \citep[e.g.][]{Owen2012,Kubyshkina2018a}.} 

An additional important length scale in the problem is the planet's ``Bondi radius'':
\begin{equation}
    R_B=\frac{GM_p}{2c_s^2}
\end{equation}
with $G$ the gravitational constant, $M_p$ the planet's mass and $c_s$ the sound speed of the gas with a temperature equal to the equilibrium temperature. This radius represents the radius at which bolometrically heated gas becomes unbound from the planet and is equivalent to the sonic radius in a bolometrically powered isothermal outflow (i.e. the Bondi radius is equivalent to the sonic radius in the core-powered mass-loss regime). {\rc In the following, we use the subscripts ``eq'' and ``pe'' to refer to quantities in the bolometrically heated and photoevaporative heated regions, respectively. The transition between these two regions occurs at $R_{\rm XUV}$. }

\begin{figure*}
    \centering
    \includegraphics[width=\textwidth]{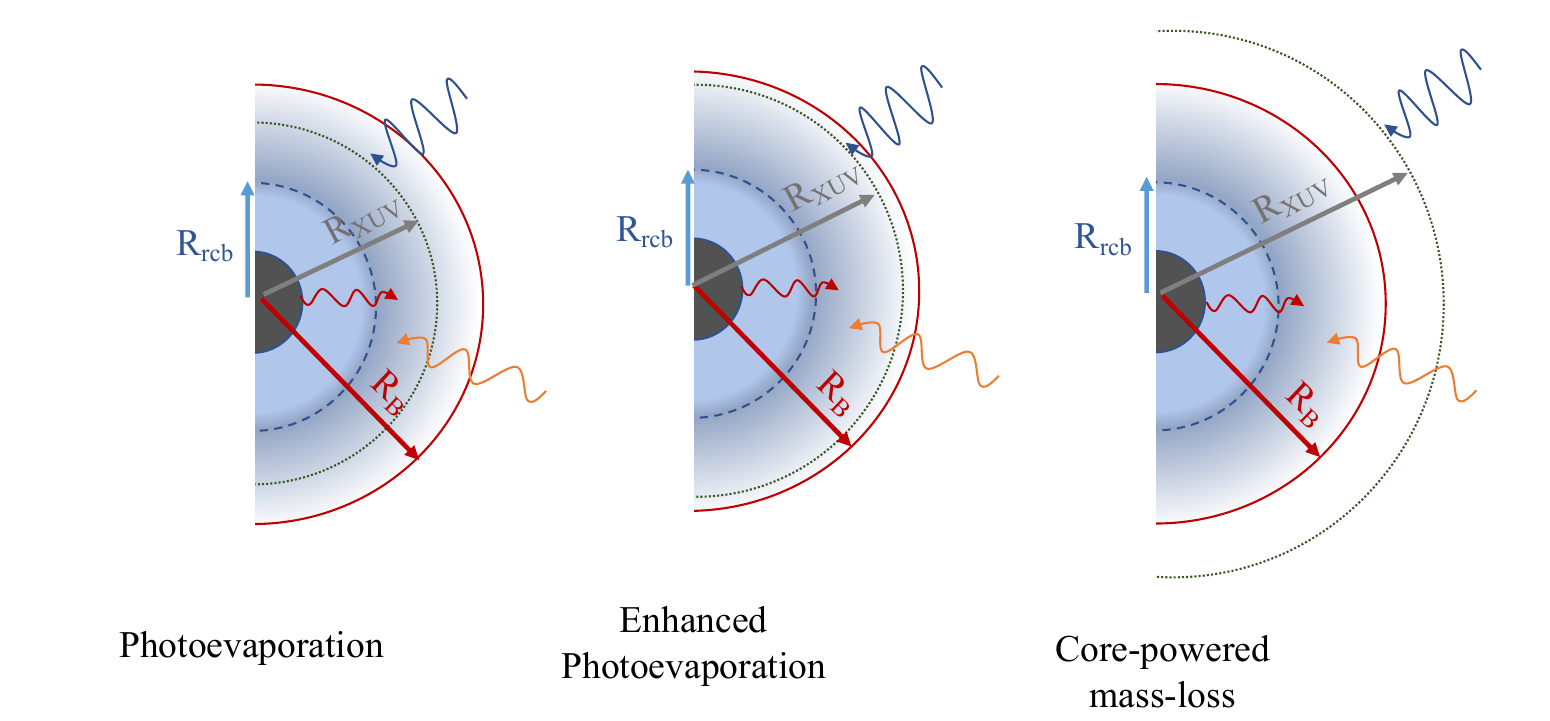}
    \caption{Schematic depicting the mass-loss regimes for ``photoevaporation'', `` enhanced photoevaporation'' and ``core-powered mass-loss''.  A planet that is undergoing 
    photoevaporation has its Bondi radius ($R_B$) located outside the region at which XUV photons can penetrate the envelope ($R_{XUV}$) such that $R_s < R_B$. Whereas a planet undergoing core-powered mass-loss has its Bondi radius inside the XUV heated region ($R_B<R_{XUV}$) and $R_s = R_B$. Even if an outflow is primarily controlled by XUV heating and $R_{XUV}<R_B$; heating from the planet's core and the bolometric luminosity of its host star can enhance the ability of a planet to absorb XUV irradiation by pushing the XUV radiation to lager heights and thus driving a more powerful photoevaporative outflow. As a result, when $R_{XUV}\gg R_p$, core-powered mass-loss and photoevaporation work in concert to drive ``enhanced'' photoevaporative outflows (middle panel). XUV photons and optical photons from the host star are shown in blue and orange, respectively. The IR/optical radiation from the planet's interior is shown in red.}
    \label{fig:schematic}
\end{figure*}

We show in the following sections that we can determine when atmospheric mass-loss transitions from core-powered to escape driven by photoevaporation by calculating the location of the Bondi radius relative to the penetration depth of XUV photons (see Figure \ref{fig:schematic}). Photoevaporation dominates the loss when the XUV penetration depth lies inside the Bondi radius ($R_{XUV}<R_B$) and core-powered mass-loss when XUV radiation is absorbed higher up in the flow ($R_B<R_{XUV}$).

In this hydrogen-dominated envelope, we approximate the mean particle mass $\mu$, as $2m_h$ (with $m_h$ the mass of the hydrogen atom) in the bolometrically heated region, $\mu=m_h$ in the XUV heated region and $\mu=m_h/2$ in any region in recombination-ionization equilibrium. {\rc To solve the problem, we approximate that these transitions are sharp, occurring at a fixed radius, while noting that simulations \citep[e.g.][]{Owen2016,Kubyshkina2018a} show these transitions can be smooth, taking place over a finite radius range. }

This work does not distinguish between core-powered mass-loss and ``boil-off/spontaneous mass-loss'' since they are both bolometrically heated outflows above the radiative-convective boundary. An evolutionary model is needed to determine the radiative-convective boundary's energy supply, allowing core-powered mass-loss and boil-off/spontaneous mass-loss to be distinguished \citep[c.f.][]{Rogers2023c}. Throughout this work, we use core-powered mass-loss to refer to this bolometrically powered outflow since we are primarily concerned with the late-time evolution of planets after disc dispersal. However, all our criteria could equally be applied to the transition between photoevaporation and boil-off/spontaneous mass-loss \citep[see,][]{Rogers2023c}.

\subsection{Core-powered mass-loss or photoevaporation?}
Using this atmosphere/outflow structure, we can now begin to consider which mass-loss mechanism dominates. More specifically, we wish to determine whether the physics of photoevaporation or core-powered mass-loss ultimately sets the mass-loss rates. 

A strongly irradiated planet that receives {\it no} XUV flux from its host star will naturally produce an approximately isothermal outflow at roughly the planet's equilibrium temperature, hence a "core-powered" outflow. Without XUV radiation, this outflow can still be shut off if it becomes too rarefied and the upper atmosphere is no longer collisional, meaning the hydrodynamic approximation is invalid. A hydrodynamic picture is applicable when the mean free path of the individual particles is smaller than the flow scale or:
\begin{equation}
    \frac{1}{n\sigma_{\rm col}} \lesssim \frac{\partial r}{\partial\log P} \label{eqn:hydro_limit}.
\end{equation}
where $n$ is the number density, $\sigma_{\rm col}$ is the collisional cross-section, $r$ is the radius from the centre of the planet and $P$ is the gas pressure. For a hydrodynamic outflow, this condition is required to be satisfied everywhere inside the sonic point. Thus, core-powered mass-loss can be shut off if the inequality in Equation~\ref{eqn:hydro_limit} becomes invalid at the sonic-point\footnote{Since density decreases with distance from the planet and the scale height increases the inequality in Equation~\ref{eqn:hydro_limit} breaks down at large distances first.}. At the sonic point ($R_s=R_B=GM_p/2c_s^2$) of an isothermal Parker wind, the flow length scale is $R_s/3$. Re-writing this inequality in terms of mass-loss rate ($\dot{M}$) we find:
\begin{equation}
    \dot{M}> \frac{12 \mu \pi c_s R_B }{\sigma_{\rm col}}.
\end{equation}
The core-powered mass-loss rate can be written as:
\begin{equation}
    \dot{M}_{\rm CP}=4\pi R_{\rm p}^2 \rho_{\rm phot} c_s \mathcal{M}(R_{\rm p}/R_{B}),
\end{equation}
where $\rho_{\rm phot}$ is the density at the planet's photosphere and $\mathcal{M}$ is the Mach number of the flow. Now, using the fact that in hydrostatic equilibrium (an adequate approximation below the sonic point) we can write the density at the photosphere to the outgoing thermal radiation as:
\begin{equation}
    \rho_{\rm phot}\approx \frac{g}{c_s^2\kappa_{\rm IR}} = \frac{2R_B}{R_p^2\kappa_{\rm IR}}, \label{eqn:phot_density}
\end{equation} where $g$ is the strength of the planet's gravitational acceleration at the photosphere and $\kappa_{\rm IR}$ is opacity to outgoing thermal, IR radiation. We then arrive at the simple condition for the case at which the outflow will just be collisional at the sonic point.:
\begin{equation}
    \mathcal{M}_{\rm phot} \gtrsim \frac{3}{2}\frac{\sigma_{\rm IR}}{\sigma_{\rm col}} \approx 5\times 10^{-10} \, \left(\frac{\kappa_{\rm IR}}{1\times10^{-2}~{\rm cm^2~g^{-1}}}\right)\left(\frac{\sigma_{\rm col}}{10^{-16}~{\rm cm^2}}\right)^{-1}.
\end{equation}
where $\sigma_{IR}$ is the absorption cross-section to IR radiation. This ``launch'' Mach number is a unique function of $R_s/R_{\rm p}$, or what's often called the ``escape parameter'' in other contexts. We find that core-powered mass-loss will shut down for $R_s/R_{\rm p} \gtrsim 14$, for canonical values of the IR opacity and collision cross-section.

However, a core-powered outflow is likely to switch to a photoevaporative if the core-powered outflow can be penetrated by XUV radiation before the sonic point. Ultimately, this means that if:
\begin{equation}
n\sigma_{\rm XUV} \frac{\partial r}{\partial\log P} \lesssim 1 \label{eqn:penetration1}
\end{equation}
at the sonic point (with $\sigma_{\rm XUV}$ the absorption cross-section to XUV radiation), the flow will likely become photoevaporative. This condition is essentially the same as in Equation~\ref{eqn:hydro_limit}, except the collision cross-section has been replaced with the cross-section to absorb ionizing photons. Since $\sigma_{\rm XUV}\sim 10^{-18}~{\rm cm^2}$ in the case of EUV photons ({\rc for atomic hydrogen}) and $\sim 10^{-21}-10^{-22}~{\rm cm^2}$ in the case of soft X-rays (for Solar metallicity gas), this means a core-powered mass-loss outflow will always be penetrated by XUV radiation interior to the sonic point before it transitions to Jeans escape. This insight is an important conclusion, as it means the breakdown of core-powered mass-loss is not primarily controlled by the outflow becoming collisionless. Thus, the transition to a collisionless Jeans escape-like outflow is more likely to occur in an XUV-heated region. This conclusion only breaks down when the XUV irradiation provides insufficient heating, as discussed in Section~\ref{sec:energy_limit}. 

Therefore, the transition between core-powered mass-loss and photoevaporation is going to primarily arise from whether XUV photons can penetrate a core-powered mass-loss outflow sufficiently deeply to affect the mass-loss rate. As discussed by \citet{Bean2021}, XUV photons can only affect the outflow if they are absorbed interior to the planet's Bondi radius, where the gas is moving sub-sonically. If they were absorbed exterior to the Bondi radius, the bolometrically heated gas would already be travelling super-sonically. Information cannot propagate upstream in a super-sonic hydrodynamic outflow, so the XUV photons would not contribute to the mass-loss rates.  Thus, a planet undergoing core-powered mass-loss has its Bondi radius residing inside the point at which XUV photons can reach. A planet undergoing photoevaporation has its Bondi radius within the XUV-heated region. We also highlight the case of ``enhanced'' photoevaporation, where the Bondi radius resides outside $R_{\rm XUV}$; however, the bolometrically heated region extends well beyond $R_p$. In this case, the bolometrically heated region allows the planet to intercept more stellar XUV photons resulting in higher mass-loss rates (and is sometimes parameterised in energy-loss models via a radius enhancement factor; e.g. \citealt{Lammer2003,Baraffe2004,Owen2019})\footnote{Some previous photoevaporation evolutionary models have implicitly included this ``enhanced photoevaporation'', whereas others have not.}.  This penetration depth argument is similar to the discussion of X-ray compared to EUV-driven photoevaporation \citep[e.g.][]{Owen2012}, and has a well-established theoretical framework in disc photoevaporation \citep[e.g.][]{Johnstone1998,Richling2000,Owen2012disc}. 


In the following sub-sections, we expand on the above discussion, and analytically derive the various transition criteria, laying the theoretical foundations for our later numerical computations. We provide a physically motivated discussion of these insights in Section~\ref{sec:first_summary}, before preceding with the numerical solutions. 

\subsubsection{Penetration by XUV radiation}\label{sec:penetration}
We have now identified one of the main reasons why core-powered mass-loss switches to a photoevaporative outflow: that the isothermal region heated by the planet's internal and stellar bolometric radiation is penetrated by XUV radiation, launching a, generally more powerful photoevaportive outflow. Thus, the criteria to switch to a photoevaporative flow due to the penetration of XUV photons is approximately:
\begin{equation}
    \int^\infty_{R_B}n_{\rm pe}(r)\sigma_{\rm XUV}{\rm d}r \lesssim 1\,, \label{eqn:optical_depth1}
\end{equation}
where $n_{\rm pe}$ is the number density in the photoevaporative region. The limiting case is determined by the point where the photoevaporative outflow is launched from the Bondi radius ($R_B$, the sonic point of the core-powered mass loss outflow)\footnote{In the following, we use $c_s$ to refer to the sound speed in the bolometrically heated region.}, in this case, the photoevaporative outflow is already trans-sonic due to its higher temperature. Thus, in this limiting case, we assume the outflow velocity outside $R_B$ is approximately constant, and the density profile falls off as $n_{\rm pe}\propto 1/r^2$. In addition, we assume that the density profile is predominately neutral near $R_B$ and ignore the effects of ionization (which we will treat later). Under these simplifications Equation~\ref{eqn:optical_depth1} becomes:
\begin{equation}
    n_{\rm pe}(R_B)\sigma_{\rm XUV}R_B = 1 \label{eqn:optical_depth2}
\end{equation}
Assuming that the photoevaporative outflow is travelling with a velocity equal to the sound speed in the XUV heated gas ($c_{\rm pe}$), then momentum balance across the transition to the photoevaporative outflow implies:
\begin{equation}
    \rho_{\rm eq}(R_B) c_s^2 \approx 2 \rho_{\rm pe}(R_B) c_{\rm pe}^2 \label{eqn:mom_balance1}
\end{equation}
{\rc where $\rho_{\rm pe}$ and $\rho_{\rm eq}$ are the mass densities in the photoevaporative and bolometrically heated regions, respectively}. It is easy to show, enforcing both mass and momentum conservation, that the bolometrically heated region is strongly sub-sonic even in the case $R_{\rm XUV}=R_B$, {\rc and follows a breeze solution}\footnote{{\rc It is important to distinguish this from the case when $R_{\rm XUV}>R_B$, where the sonic point would occur at $R_B$.}}. Thus, neglecting its momentum flux {\rc in the bolometrically heated region} allows us to relate $n_{\rm pe}$ to $\rho_{\rm phot}$ and ultimately to the planetary properties. As the bolometrically heated region is moving sub-sonically, its density profile as a function of radius ($r$) is given approximately by the hydrostatic solution:
\begin{equation}
    \rho_{\rm eq}(r) = \rho_{\rm phot}\exp\left[ \frac{2R_B}{R_p}\left(\frac{R_p}{r}-1\right)\right] \label{eqn:iso_hydrostatic}
\end{equation}
Combining Equations~\ref{eqn:optical_depth2}-\ref{eqn:iso_hydrostatic} with Equation~\ref{eqn:phot_density}, we arrive at a transcendental equation that describes the transition from core-powered mass-loss to photoevaporation:
\begin{equation}
    2 \left(\frac{R_B}{R_{\rm p}}\right)^2 \exp\left[ 2\left(1-\frac{R_B}{R_{\rm p}}\right)\right] \left(\frac{c_s}{c_{\rm pe}}\right)^2\left(\frac{\sigma_{\rm XUV}}{\sigma_{\rm IR}}\right) = 1 \label{eqn:transition1}
\end{equation}
{\rc The factor of two at the beginning of the expression arises since} we have assumed the base of the photoevaporative outflow is atomic gas and the regions around the IR photosphere are molecular, and, as such, there is a factor two change in the mean molecular weight between these two positions. Thus, we see the transition point can be described in terms of the ratio $R_B/R_{\rm p}$ (or the escape parameter). Since any bound planet will require $R_{\rm p}\ll R_B$, we can expand Equation~\ref{eqn:transition1} to find:
\begin{equation}
    \frac{R_B}{R_{\rm p}}\approx \log \left[\sqrt{2}e\left(\frac{c_s}{c_{\rm pe}}\right)\sqrt{\frac{\sigma_{\rm XUV}}{\sigma_{\rm IR}}}\right] + \mathcal{O}(1) \sim 9 + \mathcal{O}(1) \label{eqn:penetrate_solution1}
\end{equation}
where the final approximate value is determined by substituting a sound-speed ratio of $c_{\rm pe}/c_s \sim 5$, $\sigma_{\rm XUV}\sim 2\times10^{-18}~$cm$^2$ and $\kappa_{\rm IR}=10^{-2}~{\rm cm^2~g^{-1}}$. We caution that this and the following expansions are only good to of order 1\footnote{The next term in the expansion is $\mathcal{O}\log\log(c_s/c_{\rm ph}\sqrt{\sigma_{\rm XUV}/\sigma_{IR}})$, and different expansion techniques can partition different order-unity coefficients in either term, hence the $\mathcal{O}(1)$ accuracy. All our expansions come from re-writing the transcendental equation in terms of the Lambert W function, which we then expand asymptotically \citep{deBruijn:1961:AMA}.}, as evidenced by the $\sim 10-20$\% difference between using this approximation and explicitly solving Equation~\ref{eqn:transition1}. Thus, Equation~\ref{eqn:penetrate_solution1} is a rough but instructive guide, especially considering the approximations made to arrive at the result. In particular, if one would assume soft X-rays drive the photoevaporative outflow, one would find a value of $R_B/R_{\rm p}$ closer to $\sim$6 (see Section~\ref{sec:Xrays}).  However, this result does imply that core-powered mass-loss is more applicable for low-mass, puffy planets, while photoevaporation applies to higher-mass, denser planets in agreement with previous simulation results \citep[e.g.][]{Kubyshkina2018a}. 

\subsubsection{Ionization-recombination balance}\label{sec:recombination}
In the previous case, we assumed that the gas in the vicinity of $R_{\rm XUV}$ was predominately neutral. However, at sufficiently high EUV fluxes, the gas can become highly ionized and thus transparent to EUV photons. This allows EUV photons to penetrate deeper into the atmosphere. When the gas becomes highly ionized, recombination is frequent, and the XUV heated region reaches an ionization-recombination balance \citep[e.g.][]{MurrayClay2009,Owen2016}. As the recombination rate is slower than the Lyman-$\alpha$ cooling rate, this thermostats the gas to $\sim 10^4$~K. Thus, one can calculate the position of $R_{\rm XUV}$ through a Str\"omgren volume argument \citep[e.g.][]{Bertoldi1990}. Following \citep{MurrayClay2009,Owen2016}, the density at $R_{\rm XUV}$ when $R_{\rm XUV}=R_B$ (i.e. the atmosphere penetration depth to XUV photons equal the sonic radius from the hydrodynamic out-flow from core-powered mass-loss) can be found by balancing ionizations with recombinations locally, using the on-the-spot approximation:
\begin{equation}
    \frac{F_{\rm XUV}}{h\bar{\nu}}=\phi_{XUV} = \int^\infty_{R_B} \alpha_Bn_{\rm pe}^2(r){\rm d}r \label{eqn:penetration_RL}
\end{equation}
where $\phi_{\rm XUV}$ is the ionizing flux in photons per unit time per unit area, and $\alpha_B$ is the case-B recombination coefficient. The photon flux is related to the energy flux ($F_{\rm XUV}$) in terms of a representative photon energy ($h\bar{\nu}$), which we choose to be 20 eV throughout this work. Following the same steps as in Section~\ref{sec:penetration}; specifically assuming $n_{\rm pe}\propto 1/r^2$, adopting momentum balance across $R_{\rm XUV}$ (Equation~\ref{eqn:mom_balance1}) and taking the bolometrically heated region to have a hydrostatic density profile (Equation~\ref{eqn:iso_hydrostatic}) we arrive at the following criteria for core-powered mass-loss to transition to photoevaporation, assuming photoevaporation takes place in the recombination limit:
\begin{equation}
    \phi_{\rm XUV}=\frac{\alpha_B}{3\sigma_{IR}^2R_{\rm p}}\left(\frac{\mu_{\rm eq}}{\mu_{\rm pe}}\right)^2\left(\frac{c_s}{c_{\rm pe}}\right)^4\left(\frac{R_B}{R_{\rm p}}\right)^3\exp\left[4\left(1-\frac{R_B}{R_{\rm p}}\right)\right]
\end{equation}
where $\mu_{\rm pe}$ and $\mu_{\rm eq}$ are the mean-molecular weights in the photoevaporative and bolometrically heated regions, respectively. This above expression can again be approximately solved by expansion to find:
\begin{equation}
    \frac{R_B}{R_{\rm p}}\approx\frac{1}{4}\log\left[\frac{9e^{4}\alpha_B}{64\sigma_{IR}^2\phi_{\rm XUV} R_{\rm p} }\left(\frac{c_s}{c_{\rm pe}}\right)^4\left(\frac{\mu_{\rm eq}}{\mu_{\rm pe}}\right)^2\right]+\mathcal{O}(1)\sim 5 +\mathcal{O}(1) \label{eqn:ion_solution1}
\end{equation}
where the last estimate has been evaluated for $F_{\rm EUV}=10^{4}~$erg~s$^{-1}$~cm$^{-2}$, $\kappa_{\rm IR}=10^{-2}$~cm$^2$~g$^{-1}$, $\mu_{\rm pe}/\mu_{\rm eq}=1/4$, $c_{\rm pe}/c_s=5$ and $\alpha_B=2.6\times10^{-13}$~cm$^3$~s$^{-1}$. This value is slightly smaller than evaluated in the standard penetration case (Equation~\ref{eqn:penetrate_solution1}) and can be understood in terms of the increased ability of EUV photons to penetrate into the atmosphere at high fluxes since they can ionize the gas, resulting in a longer photon mean-free path. However, like in Equation~\ref{eqn:penetrate_solution1}, the key result remains the same: the transition between photoevaporation and core-powered mass-loss occurs approximately at constant $R_p/R_B$, although in this case, there is an explicit, albeit logarithmic dependence on the planet's radius. 

\subsubsection{Heating Limitation}\label{sec:energy_limit}
The previous analysis in Section~\ref{sec:penetration} and \ref{sec:recombination} {\it assumes} that the ionizing radiation provides sufficient heating to drive a powerful flow. However, in the limit of a low ionizing flux, it may not provide any additional heating. This example is shown by \citet{Schulik2023_new}, who found a smooth transition from a core-powered outflow to a photoevaporative outflow as the ionizing flux increased at fixed bolometric flux. 

We consider this ability to switch from core-powered mass-loss to photoevaporation to be a heating requirement. Namely, that the high-energy field has sufficient energy to drive a more powerful outflow than core-powered mass-loss. Thus, the transition occurs when the mass-loss rate provided by core-powered mass-loss is comparable to photoevaporation. To explore this transition approximately, we assume the photoevaporation rate is given by the commonly used ``energy-limited'' model \citep[e.g.][]{Baraffe2004}, where:
\begin{equation}
    \dot{M}_{\rm pe}=\eta F_{\rm XUV} \frac{\pi R_{XUV}^3}{4GM_p} \label{eqn:EL_mdot}
\end{equation}
where $\eta$ is the mass-loss efficiency. {\rc Our choice of the factor of 4 in the numerator accounts for the fact planets absorb flux over an area of $\pi R_{\rm XUV}^2$, but mass-loss can occur over the full sphere, i.e. $4\pi R_{\rm XUV}^2$, \citep{Owen2019}.}  Thus, equating this mass-loss rate to that given by core-powered mass-loss $\dot{M}_{\rm CP}$, we find the transition to photoevaporation occurs at a critical flux of:
\begin{equation}
    F_{\rm XUV} =\frac{32 GM_pR_Bc_s}{\eta R_{\rm XUV}^3\kappa_{\rm IR}}\mathcal{M}\left(R_{\rm p}/R_B\right) \label{eqn:energy_limit}
\end{equation}
where $R_{\rm XUV}$ is found by determining:
\begin{equation}
    \int^\infty_{R_{\rm XUV}} n(r)\sigma_{\rm XUV}{\rm d}r = 1. \label{eqn:tau_1_energy}
\end{equation}
Now in the limiting case that the XUV irradiation provides insufficient heating the density profile $n(r)$ will simply be that for the Parker wind solution for the core-powered mass-loss outflow. For the heating limit to even be relevant $R_{\rm XUV}<R_B$; thus, we can approximate the density profile $n(r)$ with the hydrostatic solution (Equation~\ref{eqn:iso_hydrostatic}). In Appendix~\ref{app:opt_depth_calc}, we show that Equation~\ref{eqn:tau_1_energy} has two limiting solutions, one for $R_p/R_B\log(\sqrt{\sigma_{XUV}/\sigma_{IR}}) \ll 1$ 
\begin{equation}
    \frac{R_{\rm XUV}}{R_p}\approx1 + \frac{R_p}{R_B}\log\left(\sqrt{\frac{\sigma_{XUV}}{\sigma_{IR}}}\right)
\end{equation}
corresponding to a dense planet, where the XUV irradiation penetrates close to $R_p$. The other limiting solution, in the case $R_p/R_B\log(\sqrt{\sigma_{XUV}/\sigma_{IR}}) \gg 1$ is:
\begin{equation}
    \frac{R_{\rm XUV}}{R_p}\approx \sqrt{\frac{\sigma_{XUV}}{\sigma_{IR}}} \exp\left(-\frac{R_B}{R_p}\right)\label{eqn:pen_energy_fluffy}
\end{equation}
Corresponding to a puffier planet where the XUV irradiation is absorbed at several planetary radii. Given typical values of the cross sections, the transition between the two solutions occurs roughly at $R_B/R_p\sim 10$. This transition occurs before the penetration criteria given in either Equation~\ref{eqn:penetrate_solution1} or \ref{eqn:ion_solution1}. Taking $R_p/R_B \ll 1$ (for both cases), $\mathcal{M}(R_p/R_B)$ becomes \citep[e.g.][]{Lamers1999}:
\begin{equation}
    \mathcal{M}(R_p/R_B)\approx \left(\frac{R_p}{R_B}\right)^{-2}\exp\left(-\frac{2R_B}{R_p}\right).
\end{equation}
Thus, the solution to Equation~\ref{eqn:energy_limit} for the dense planet, with $R_{\rm XUV}\approx R_p$ is:
\begin{equation}
    \frac{R_B}{R_p}\approx \frac{1}{2}\log\left(\frac{108 g c_s}{\eta\kappa_{IR}F_{\rm XUV}}\right) + \mathcal{O}(1) 
\end{equation}
or for the puffier planet with $R_{\rm XUV}$ given by Equation~\ref{eqn:pen_energy_fluffy} is:
\begin{equation}
    \frac{R_B}{R_p}\approx \log\left[\frac{\eta\kappa_{IR}F_{\rm XUV}}{864 g c_s}\left(\frac{\sigma_{XUV}}{\sigma_{IR}}\right)^{3/2}\right] + \mathcal{O}(1) 
\end{equation}
where, like in the previous sections, these approximate solutions have been obtained by expansion. Although, we caution that for high flux values, this ``heating limit'' yields no solution and photoevaporation can occur for any planet (see Section~\ref{sec:numerical}). Thus, at sufficiently low XUV irradiation levels, we expect this energy limit to push the transition from core-powered mass loss to photoevaporation to denser planets. In our numerical evaluations, we find scenarios where, as a planet loses mass (and shrinks), it will transition from core-powered mass-loss to photoevaporation (due to XUV penetration), but as the XUV flux drops, it can transition back to core-powered mass-loss (see track ``A'' in Figure ~\ref{fig:XvsM}), before becoming photoevaporative again when the planet's atmosphere becomes thinner. 

\subsection{Summary}\label{sec:first_summary}
By considering what physical processes determine whether the outflow is predominantly powered by XUV heating or by bolometric heating from the interior and star, we have determined the basic criteria for which each mass-loss mechanism controls the outflow properties. The key result is that the transition is primarily controlled by the ratio of the Bondi radius to the planet's radius, with typical values in the range of 6-11. This result is in agreement with the simulations of \citet{Kubyshkina2018a}, that found the transition was best described in terms of the ``escape parameter'' (which is the same as $R_B$ apart from an order-unity multiplicative factor).  All other properties give rise to a slowly varying logarithmic dependence. The fundamental reason is that all criteria depend, either explicitly or implicitly, on XUV photons' ability to penetrate the approximately isothermal bolometrically heated atmosphere. The scale height of such an atmosphere depends only on $R_B/R_p$ (Equation~\ref{eqn:iso_hydrostatic}) and is exponential. Hence, any optical depth into such an atmosphere will naturally depend directly on $R_B/R_p$ but logarithmically on other parameters. The logarithmic sensitivity arises for the same reason that forming planets are only logarithmically sensitive to the disc conditions \citep[e.g.][]{Piso2015}.

We have identified that the primary transition criteria is the ability of XUV photons to penetrate the interior to the Bondi radius ($R_B$, the core-powered mass-loss sonic point), providing additional heating and hence higher mass-loss rates. Generally, larger planets (bigger $R_p/R_B$) have core-powered mass-loss outflows, while smaller planets (smaller $R_p/R_B$) have photoevaporative outflows. This transition can either occur for an energy-limited or recombination-limited outflow, with recombination-limited outflows becoming photoevaporative for larger planetary radii. This is because of their ability to reach high ionization fractions, reducing the optical depth to XUV photons allowing them to penetrate deeper. At low XUV fluxes, XUV photons may be able to penetrate the outflow, but they do not provide additional heating, and the outflow can remain driven by core-powered mass-loss.

Finally, our analysis has indicated that even if an outflow is primarily controlled by XUV heating (and hence ``photoevaporative''), bolometric heating from the core and star is not unimportant. Ultimately, it's this heating source that provides the energy to lift fluid parcels from the radiative-convective boundary up to $R_{\rm XUV}$\footnote{It's important to note the fundamental difference between this scenario for highly irradiated planets, where bolometric heating provides this energy and the original formalism of ``energy-limited'' mass-loss \citet{Watson1981,Lammer2003}, where conduction of XUV irradiation provides this energy.}. This bolometric heating can push the XUV absorption to higher heights, enhancing the ability of the planet to absorb XUV irradiation and driving a more powerful photoevaporative outflow. Thus, core-powered mass-loss and photoevaporation can work in concert to drive ``enhanced'' photoevaporative outflows, especially for planets that have just transitioned from core-powered mass-loss to photoevaporation as they cool and lose mass.

\section{Approximate numerical solutions} \label{sec:numerical}
Full radiation-hydrodynamic simulations that include the radiation from the planet's interior, the bolometric radiation from the star, and the stellar ionizing radiation are required to fully map out the parameter space. However, we can improve our analytic approach by numerically relaxing some of the assumptions. In addition, we can assess the role the bolometrically heated layer plays in enhancing photoevaporation. Specifically, in Section~\ref{sec:penetration}, our solution depends on the unknown sound speed in the XUV heated region for an ``energy-limited'' photoevaporative outflow.  

To progress, we still assume an isothermal outflow for the photoevaporative region, but with a sound speed, we numerically obtain. In this simplification, we assume that the launch velocity at $R_{\rm XUV}$ is either the one given by the trans-sonic Parker wind solution or the sound speed (whichever is smaller; this prevents unphysical super-sonic launching of the wind). This requires using a generalised Parker wind model, described in Appendix~\ref{app:general_parker}.

We then numerically integrate the photoevaporatively heated outflow's density profile to calculate the optical depth to XUV photons (i.e. we numerically solve Equation~\ref{eqn:optical_depth1}). For a given $R_{\rm XUV}$, there is then a family of solutions, each with a different sound speed and hence different mass-loss rate, that satisfies the criteria that the optical depth to XUV photons throughout the photoevaporative region is unity. 

Thus, we solve for the appropriate sound speed and hence $R_{\rm XUV}$ to match the energy-limited model's mass-loss rate with an efficiency of 0.1 (Equation~\ref{eqn:EL_mdot}). If the photoevaporative outflow temperature we find is below the planet's equilibrium temperature, we identify the outflow as core-powered mass-loss because, while the XUV photons can penetrate, they do not provide additional heating (the ``heating limit'' described in Section~\ref{sec:energy_limit}) and therefore don't enhance the outflow. It is well known that above a temperature of $10^4$~K,  Lyman-$\alpha$ cooling dominates, and the outflow is no longer energy-limited \citep[e.g.][]{MurrayClay2009,Owen2016}. To mimic this effect, we do the following: If matching the energy-limited model requires a temperature in excess of $10^4$~K, we fix the outflow's temperature to be $10^4$~K and reduce the mass-loss rate below the energy-limited value. Furthermore, recombination can become important once the gas temperature has reached $10^4$~K. If the time scale for a proton to recombine becomes shorter than the flow time scale, the outflow enters radiation-recombination balance \citep[e.g.][]{Bear2011} and the mass-loss has a square-root dependence on XUV flux \citep[e.g.][]{MurrayClay2009}. Thus, for outflows with temperatures of $10^4$~K, we compare the recombination time to the flow timescale at $R_{\rm XUV}$. If the recombination time is shorter than the flow time scale, we switch to using recombination-limited outflows. Thus, instead of solving Equation~\ref{eqn:optical_depth1}, we numerically solve Equation~\ref{eqn:penetration_RL}.

Finally, to make a connection to a real planetary structure (i.e. one with a specific photospheric radius or envelope mass), we then solve for the value of $R_{\rm XUV}$ such that there is momentum balance across the transition from the bolometric heated region to the photoevaporative region. We assume, as previously, that the opacity to outgoing thermal irradiation is $\kappa_{\rm IR}=10^{-2}$~cm$^2$~g$^{-1}$. Since the bolometrically heated region has to be sub-sonic before the transition into the photoevaporative region (for the outflow to be identified as photoevaporation-dominated), we neglect the momentum-flux in the bolometric region and only consider the contribution from thermal pressure (as typically done in models of external disc photoevaporation -- \citealt{Johnstone1998,Owen2021,Owen2023}). Namely the matching criteria to solve for $R_{\rm XUV}$ is:
\begin{equation}
    \rho_{\rm eq}(R_{\rm XUV})c_s^2 = \rho_{\rm pe}(R_{\rm XUV})\left(u_{\rm pe}(R_{\rm XUV})^2+c_{\rm pe}^2\right). \label{eqn:momentum_balance}
\end{equation}
To solve this root-finding problem, we use the {\sc brentq} method provided in {\sc scipy} \citep{2020SciPy-NMeth}, both for the sound-speed and value of $R_{\rm XUV}$, using a relative tolerance of $10^{-13}$. The optical depth through the outflow is computed through numerical integration using the trapezoidal method on a logarithmically spaced grid between $R_{\rm XUV}$ and five times the maximum value of either $R_s$ or $R_{\rm XUV}$ on 250 cells, assuming the optical depth at the outer boundary is zero. Since the photospheric radius does not exactly correspond to the planetary radius that an optical transit observation would measure, we also compute the planet's transit radius through direct numerical integration of our density profile, assuming it to be spherically symmetric, using an optical opacity of $\kappa_{\rm op}=4\times10^{-3}$~cm$^2$~g$^{-1}$ \citep[e.g.][]{Guillot2010}. This numerical integration is performed using the adaptive Gauss-Kronrod quadrature method in {\sc quadpack}, with the python interface provided by {\sc scipy}, to a relative tolerance of $10^{-8}$. 

We do not enforce mass conservation across the interface. This is because the photoevaporative outflow could be so powerful that the bolometrically heated layer cannot supply the required mass-loss rate (i.e. remaining sub-sonic while satisfying Equation~\ref{eqn:momentum_balance}). When this occurs, the photoevaporative outflow will slowly ``eat'' into the bolometrically heated layer, pushing $R_{\rm XUV}$ to smaller values (we refer to this as ``ravenous'' photoevaporation). This is conceptually similar to the transition between expanding R-type and stationary D-type ionization fronts around massive stars \citep[e.g.][]{SpitzerBook}. We check all our solutions for any occurrence of ravenous photoevaporation. We do not find any examples in the parameter space explored in this work, though this does not mean it never occurs in planetary mass loss. 

\subsection{Results}

\begin{figure*}
    \centering
    \includegraphics[width=\columnwidth]{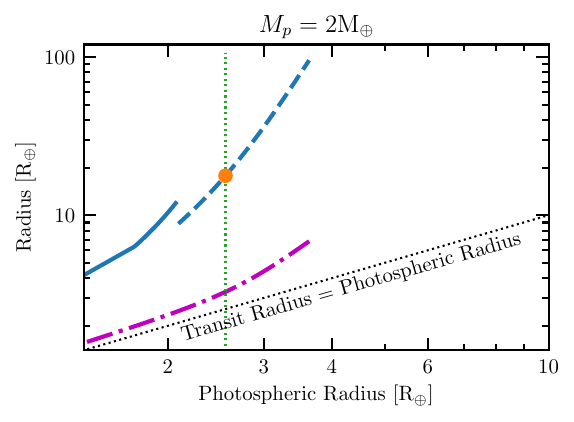}
    \includegraphics[width=\columnwidth]{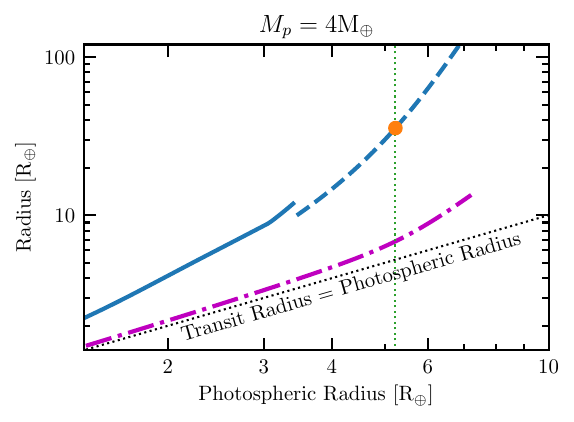}
    \includegraphics[width=\columnwidth]{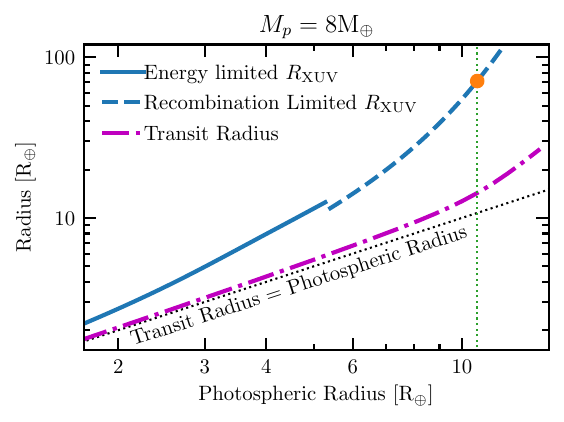}
    \includegraphics[width=\columnwidth]{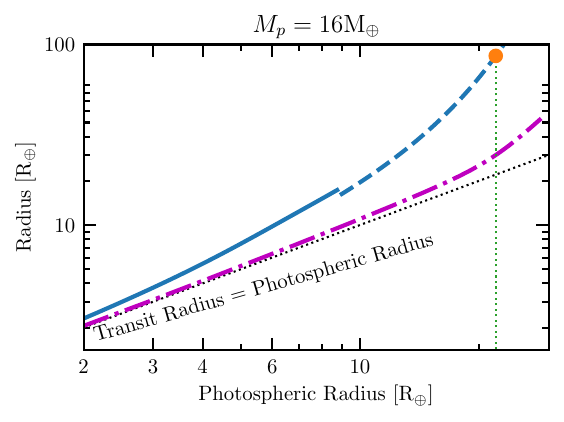}
    \caption{The radius to which XUV photons penetrate as a function of the planet's photospheric radius for planets of various masses, an equilibrium temperature of 1000~K and $L_{\rm XUV}/L_{\rm bol}=10^{-4}$. The transition between the solid and dashed lines shows where photoevaporation transitions from being energy limited to recombination limited and the XUV photons can penetrate deeper into the planet's atmosphere (e.g. at a photospheric radius of $\sim 5.2$~R$_\oplus$ for the 4~M$_\oplus$ case). The orange point shows the transition between photoevaporation and core-powered mass loss and the green dotted line indicates the photospheric radius when this transition happens. The purple dashed line shows the optical transit radius as a function of the planet's photospheric radius.}
    \label{fig:example_radius}
\end{figure*}

An example result of our calculations is shown in Figure~\ref{fig:example_radius} where we show the radius of the XUV penetration depth as a function of a planet's photospheric radius for planets of various masses, an equilibrium temperature of 1000~K and the ratio of the XUV to the bolometric flux of the star of $10^{-4}$. 
The evolution of the characteristic radii shown in Figure~\ref{fig:example_radius} demonstrates the typical evolution across the parameter space. For fixed planet mass, as the photospheric radius is increased (effectively increasing atmosphere mass fraction or decreasing a planet's age), the radius of XUV penetration increases; for small dense planets this typically begins in the ``energy-limited'' regime \footnote{We note for very high-density planets, the outflow timescale will become longer than the recombination timescales and return to a recombination limited outflow \citep{Owen2016}.}; however, once the planet's gravity becomes too weak, the outflow transitions to recombination limited \citep[e.g.][]{Owen2016}. The sharp, small drop in $R_{\rm XUV}$ arises from the assumption that for energy-limited outflows, we ignore recombination photons, which can penetrate and ionize the planet's atmosphere, decreasing $R_{\rm XUV}$, whereas in the recombination limited case, they are fully accounted for. In reality, as one approaches the transition, the energy-limited $R_{\rm XUV}$ would smoothly attach to the recombination-limited case. Eventually, with increasing photospheric radius, the XUV penetration depth will exceed the sonic point of the core-powered mass-loss outflow (shown by the orange point in Figure~\ref{fig:example_radius}), and the outflow transitions from photoevaporation at small photospheric radii, to core-powered mass-loss at large photospheric radius. As the planet becomes less dense, the transit radius adds a non-negligible correction to the photospheric radius. At the point of transition from core-powered mass-loss to photoevaporation, it is tens of percent larger. 

\begin{figure*}
\centering
\includegraphics[width=\textwidth]{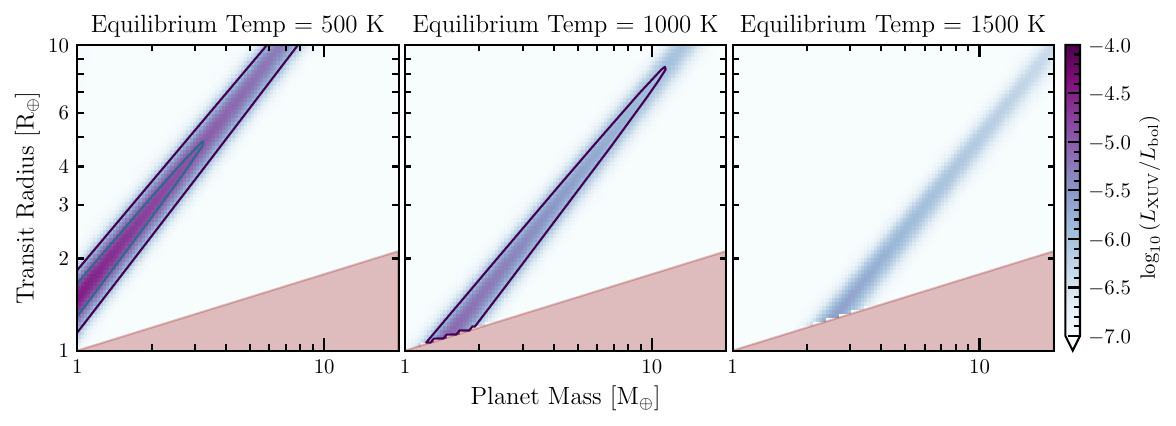}
\caption{The {\rc colour map shows the threshold value of $L_{\rm XUV}/L_{\rm bol}$ below which the outflow transitions to core-powered mass-loss from photoevaporation due to insufficient XUV heating} for different planetary equilibrium temperatures. The shaded brown regions represent Earth-like ``rocky'' cores with no atmosphere. The contours show $L_{\rm XUV}/L_{\rm bol}$ at values of $10^{-6}$, $10^{-5}$. Typically, there are either two radii at which the transition occurs at a given mass or none, as discussed in Section~\ref{sec:energy_limit}. Given that typical values of $L_{\rm XUV}/L_{\rm bol}$ are in the range of $10^{-6}$ to $10^{-3}$ for most late-type stars, this heating limit only applies in a narrow region of parameter space at cool equilibrium temperatures.} \label{fig:energy}
\end{figure*}

Figure~\ref{fig:example_radius} also indicates that, as the photospheric radius increases, the XUV penetration depth is pushed to ever larger radii in a super-linear fashion. This means that while the physics of photoevaporation ultimately controls the mass-loss rates, they are enhanced by core-powered mass-loss above the value that would be found purely from using the photospheric or transit radius. As discussed above, this ``enhanced'' photoevaporation is ultimately driven by the energy input in the isothermal layer from both the stellar irradiation and the planet's cooling luminosity, which supplies material to a larger XUV penetration depth that can absorb a higher number of XUV photons. 

We can now perform our calculation over a range of different planet masses, equilibrium temperatures and ratios of bolometric to XUV flux to determine the general conditions for the transition from photoevaporation to core-powered mass loss. In Figure~\ref{fig:energy}, we show the ratio of $XUV$ luminosity to bolometric luminosity below which the XUV irradiation would provide insufficient heating to overpower the core-powered mass-loss outflow. Thus, even if XUV photons can penetrate inside the core-powered mass-loss sonic point, the outflow will still be core-powered mass-loss below this critical value of $L_{\rm XUV}/L_{\rm bol}$. As expected from our discussion in Section~\ref{sec:energy_limit}, we find two radii (at fixed planet mass) where one would transition from photoevaporation to core-powered mass-loss and back. However, the critical values of $L_{\rm XUV}/L_{\rm bol}$ only reach observed values for typical late-type stars for cooler equilibrium temperatures. Thus, this heating limit is not the main transitioning criterion, although it will apply to the important case of temperate, low-mass planets, such as those identified in short-period orbits around M-dwarfs. 

\begin{figure}
    \centering
    \includegraphics[width=\columnwidth]{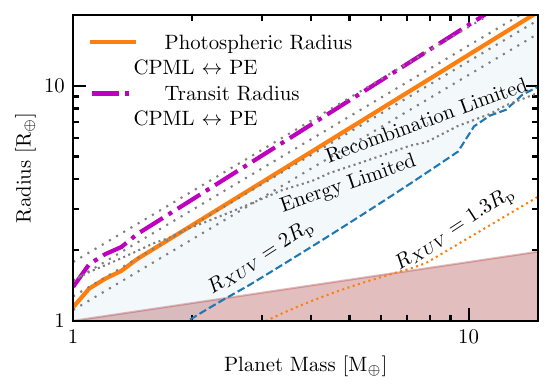}
    \caption{The various mass-loss regimes as a function of planet mass and radius for an equilibrium temperature of 1000~K and $L_{\rm XUV}/L_{\rm bol}=10^{-4}$. The transition between core-powered mass-loss (at large radii) and photoevaporation at small radii typically occurs at fixed $R_p/R_B$. The wide-spaced dotted lines show contours of constant $R_p/R_B$ for 1/5, 1/6, 1/7 \& 1/8. The {\rc narrow-dotted} line shows the transition from recombination-limited to energy-limited photoevaporation, indicating core-powered mass-loss transitions to recombination-limited photoevaporation for these parameters. The {\rc wide} dotted lines show lines of constant $R_{\rm XUV}/R_p$, where planets above the blue {\rc dashed} line with $R_{\rm XUV}=2R_p$ are representative of those undergoing ``enhanced'' photoevaporation (highlighted by the shaded blue region). The kink in this line occurs when photoevaporation transitions from energy-limited to recombination-limited. The shaded brown regions represent Earth-like ``rocky'' cores with no atmosphere. }
    \label{fig:Single_model}
\end{figure}

\begin{figure*}
    \centering
    \includegraphics[width=\textwidth]{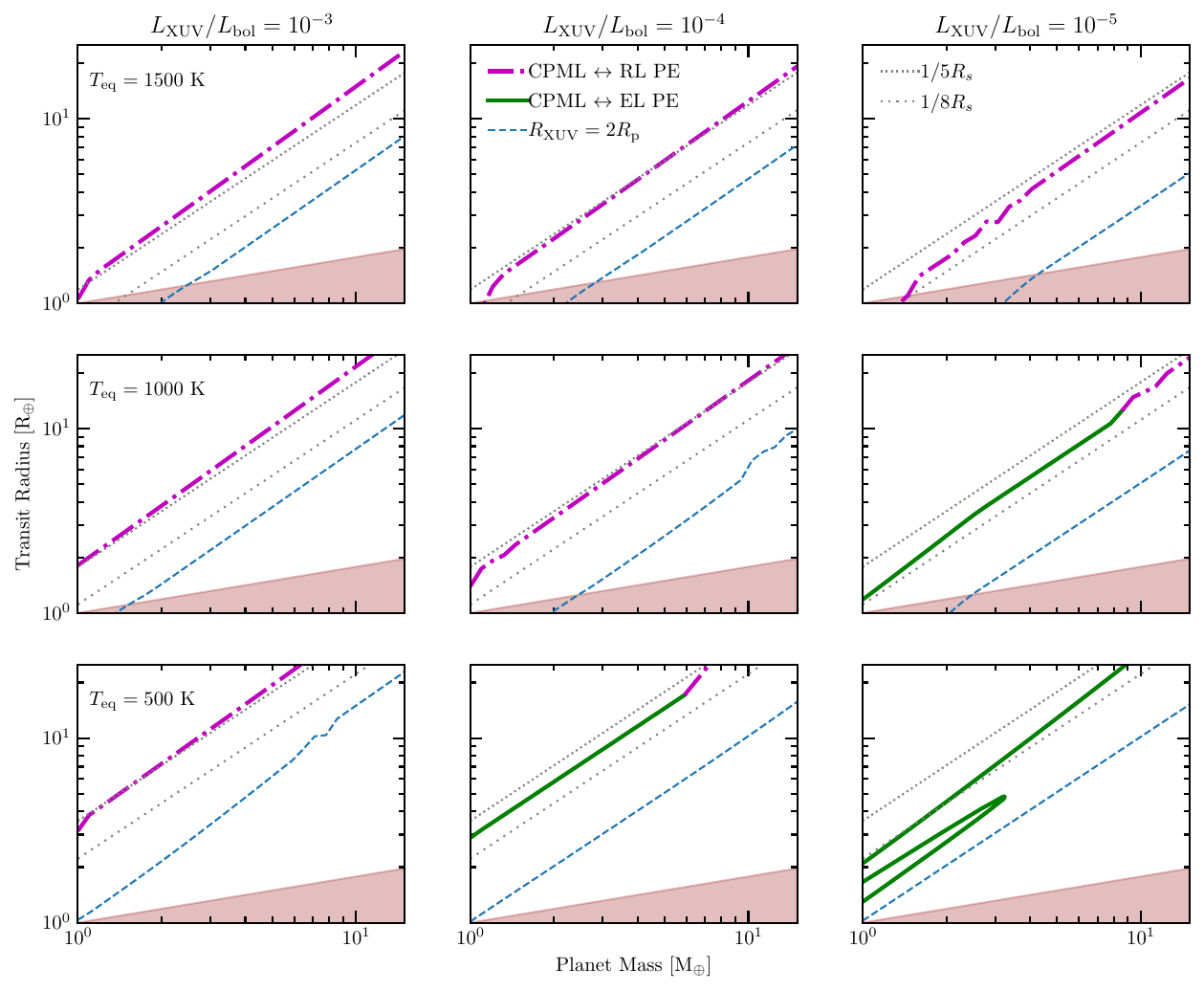}
    \caption{The transit radius at which core-powered mass-loss (at large radii) transitions to photoevaporation (at small radii) for different equilibrium temperatures and XUV fluxes. This transition occurs to either recombination-limited photoevaporation (magenta, dash-dotted line) or energy-limited photoevaporation (green, solid-line). The grey, dotted lines show lines of constant fractions of $R_B$, and the dashed line shows the region above, which we consider photoevaporation to be ``enhanced''. The kinks in these lines occur when photoevaporation transitions from energy-limited to recombination-limited. The shaded brown regions represent Earth-like ``rocky'' cores with no atmosphere.  }
    \label{fig:parameter_study}
\end{figure*}

Having investigated the heating limit transition, we now explore the penetration limit and the role of ``enhanced'' photoevaporation. In Figure~\ref{fig:Single_model}, we show the various mass-loss regimes as a function of planet mass and radius for $L_{\rm XUV}/L_{\rm bol}=10^{-4}$ and equilibrium temperature 1000~K. As expected, the transition occurs at roughly a fixed value of $R_p/R_B$, where the slight increase can be explained in terms of the logarithmic dependence of $R_p/R_B$ on $R_p$ from Equation~\ref{eqn:ion_solution1}. For these values of the XUV and bolometric flux, the transition between core-powered mass-loss and photoevaporation occurs mainly in the recombination limit (a result that appears to hold over most of the parameter space: see Figure~\ref{fig:parameter_study}). The correction between the transit radius and photospheric radius is a small but important correction increasing the transition radii by several 10s of percent. Finally, the region of enhanced photoevaporation, which we take to mean where $R_{\rm XUV}\sim 2 R_p$, encompasses a significant region of parameter space. 

We now expand our range of parameters to roughly cover the range of XUV fluxes and equilibrium temperatures covered by close-in, low-mass exoplanets. The result for the transit radius at which core-powered mass-loss transitions to photoevaporation and the range of parameters in which photoevaporation is ``enhanced'' is shown in Figure~\ref{fig:parameter_study}.  
As expected, the transition occurs at roughly fixed $R_p/R_B$ with the value ranging between $1/5$ and $1/8$. The slow changes in these values mirror the logarithmic dependence on XUV flux and sound speed in the bolometric region found in Equations~\ref{eqn:penetrate_solution1} and \ref{eqn:ion_solution1}. This plot also confirms the heating limit only matters in a small region of parameter space. In most of the parameter space, core-powered mass-loss will transition to recombination-limited photoevaporation, but photoevaporation becomes energy-limited everywhere at low XUV irradiation levels. 


\section{Discussion}
In the absence of XUV irradiation, highly irradiated planets will undergo hydrodynamic mass-loss in the form of an isothermal wind with a temperature of order the equilibrium temperature. This core-powered mass-loss outflow gets its energy from the core and envelope's cooling luminosity as well as from the star's bolometric output. We have shown that under conditions typical of close-in, low-mass planets XUV photons can penetrate this core-powered mass-loss outflow, providing extra heating before it becomes collisionless. This is because the collision cross-section is significantly larger than the cross-section to absorb XUV photons. Thus, the breakdown of the hydrodynamic limit is not really a concern in the case of core-powered mass loss. 

We have shown that the primary controlling physics determining whether an outflow is photoevaporative or core-powered mass-loss is the ability of XUV photons to penetrate interior to core-powered mass-loss's sonic point ($R_B$). This results in the transitioning criteria occurring at roughly a constant value of the planet's radius to its Bondi radius ($R_p/R_B$), where the exact value depends logarithmically on planetary and stellar properties. In general, photoevaporation will be operating in the recombination limited regime when the transition from core-powered mass-loss to photoevaporation occurs; this is because the planet's gravity is weak at $R_B$ by construction, resulting in a large flow length scale (and hence a long flow time scale) allowing sufficient time for protons to recombine \citep[e.g.][]{Owen2016}. However, at low XUV irradiation levels and cool equilibrium temperatures, the transition will occur to either energy-limited photoevaporation as the ionization rate is insufficient to reach ionization-recombination equilibrium, or the outflow can remain in the core-powered mass-loss regime for smaller planets due to insufficient XUV heating. 

We have also mapped out  ``enhanced'' photoevaporation, where, while photoevaporation sets the mass-loss rate, a sub-sonic core-powered mass-loss outflow resupplies the XUV heated region. In this case, the sub-sonic core-powered mass-loss maintains an XUV absorption radius far enough from the radiative-convective boundary allowing the planet to absorb more of the star's XUV output. Since most photoevaporation occurs early when the star's $L_{\rm XUV}/L_{\rm bol}$ is $\sim 10^{-3}$, the most common young planet with a core mass of $\sim 4-5$~M$_\oplus$ and radius of $2.5-3$~R$_\oplus$ at an equilibrium temperature of $\sim 1000$~K \citep[e.g.][]{Rogers2021}, will undergo photoevaporation around the boundary of this ``enhanced photoevaporation'' region (Figure~\ref{fig:parameter_study}). 

\subsection{Relationship to planetary properties}
\begin{figure*}
    \centering
    \includegraphics[width=\textwidth]{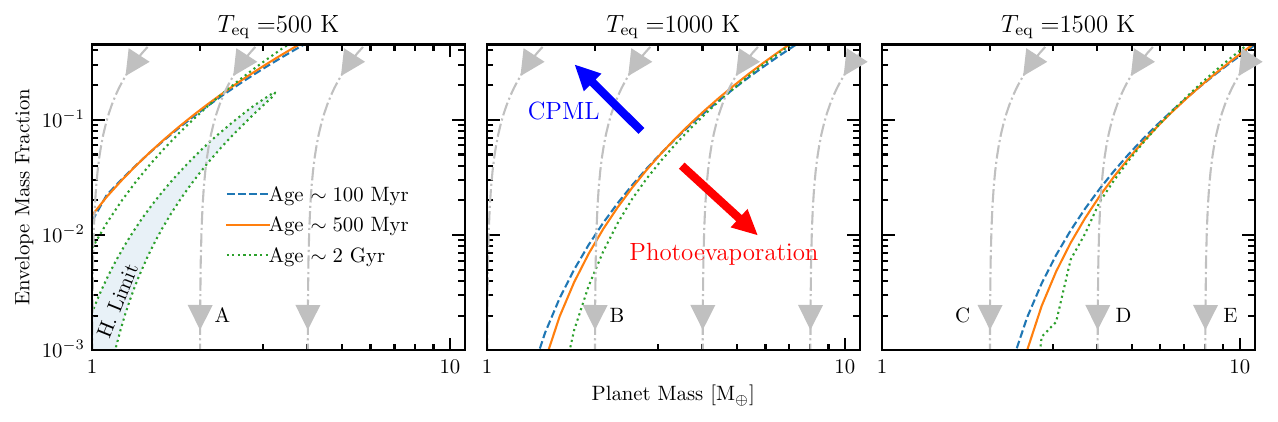}
    \caption{The hydrogen-dominated envelope mass fraction at which core-powered mass-loss (at large envelope mass fractions) transitions to photoevaporation at (small envelope mass fractions), shown for different ages (100~Myrs=blue-dashed, 500~Myrs=yellow-solid and 2~Gyrs=green-dotted lines), {\rc assuming a Solar-mass star}. The three panels correspond to equilibrium temperatures of 500K, 1000K and 1500K from left to right, respectively. Atmospheric-mass loss is dominated by core-powered mass-loss for lower-mas planets with larger-envelope mass fractions and transitions to photoevaporation-dominated for larger-mass planets and lower-envelope mass-fractions as shown by the blue, yellow and green lines and indicated by the red and blue errors in the middle panel. The exact location of the transition between these two regimes depends on the equilibrium temperature, as can be seen by comparing the results shown in the three panels. The shaded region bounded by the green dashed line represents the region where XUV photons can penetrate the core-powered mass-loss outflow but provide insufficient heating and arising from the $L_{\rm XUV}/L_{\rm bol}=10^{-5}$ contour in Figure~\ref{fig:energy} for an equilibrium temperature of 500~K. The grey, dot-dashed lines with the arrows indicate the trajectory of planets with core masses of 1, 2, 4 \& 8 M$_\oplus$. These trajectories display the rich diversity in mass-loss regimes and histories; illustrative examples are labelled ``A'' through ``E''. In general, the planet mass-loss trajectories indicate that a planet can transition from core-powered mass-loss to photoevaporation but not vice-versa. However, trajectories ``A'' and ``B'' indicate special cases discussed in the text. Trajectory ``D'' represents an example where atmospheric loss transitions from core-powered mass-loss to photoevaporation, and trajectories ``C'' and ``E'' show cases where a planet's mass loss is dominated entirely by core-powered mass-loss (C) or photoevaporation (E), respectively.}
    \label{fig:XvsM}
\end{figure*}
In our previous discussion, we have worked analytically in terms of the planet's photosphere to outgoing thermal IR radiation (which we call the planet's radius, $R_p$) as it is well defined. Numerically, we have also determined the planet's optical transit radius, as this is the observed quantity. While similar, the transit radius is always 10s of percent larger than $R_p$ at the transition boundary from core-powered mass-loss to photoevaporation. However, neither radii encapsulate the fundamental structure of the planet's envelope. The planet's radiative-convective boundary sets the transition between the adiabatic interior and the potentially outflowing atmosphere. For dense planets, this radius is similar to its photospheric and transit radius; however, for puffier planets, it can be different, and it is a time-evolving quantity. Therefore, to relate our results to planetary structure, we express the transition between core-powered mass-loss and photoevaporation in terms of the envelope mass fraction. 

We use {\sc mesa} models \citep{Paxton2011,Paxton2013,Paxton2015} {\rc to compute the relationship between photospheric radius, envelope mass fraction and age. We do this by evolving a grid of models at fixed envelope mass fraction.} The models are setup in an identical way to those described in \citep{Owen2020} for Earth-like core compositions. Our ratios of XUV to bolometric fluxes are converted to planetary ages using the empirical relations of \citet{Rogers2023} for a Solar mass star. Where values of $L_{\rm XUV}/L_{\rm bol}$ of $10^{-3}$, $10^{-4}$ and $10^{-5}$ correspond to ages of approximately, 100 Myr, 500 Myr and 2 Gyr. The envelope mass fractions at which core-powered mass-loss transitions to photoevaporation are shown in Figure~\ref{fig:XvsM}, where core-powered mass-loss dominates for lower-mass planets with comparatively larger envelope mass fractions (i.e. upper-right area in figure 7), and photoevaporation dominates for higher-mass planets with comparatively lower envelope mass fractions (i.e. lower-left area in figure 7). If the typical close-in planet is born with a few-percent envelope mass fraction around a $4$~M$_\oplus$ core, then we see that for equilibrium temperatures $\gtrsim 1500$~K core-powered mass-loss dominates the bulk of the mass loss, while for cooler planet's photoevaporation dominates the bulk of the mass-loss. 

Figure~\ref{fig:XvsM} allows us to assess the mass-loss histories of various planets. The grey, dot-dashed lines show the trajectories of planets with 1, 2, 4 \& 8 M$_\oplus$ cores. These trajectories are essentially vertical because the envelope mass is only a small fraction of the planet's total mass, resulting in limited curvature at envelope mass fractions above $10\%$. The plotted trajectories cross the mass-loss transition (moving from higher envelope mass fraction to lower envelope mass fractions) from core-powered mass-loss to photoevaporation. Thus, there are three typical planetary pathways: (i) a hot, low-mass planet (e.g. the $T_{\rm eq}=1500~$K, 2~M$_\oplus$ core will start in the core-powered mass-loss regime and is expected to continue in this mass-loss regime until it is stripped of its envelope (e.g. track C in figure~\ref{fig:XvsM}). (ii) For the same equilibrium temperature (i.e., $T_{\rm eq}=1500~$K), a planet with a core mass $\sim 4~$M$_\oplus$ born with an envelope mass fraction of a few per cent or more will start off undergoing core-powered mass-loss, but as its envelope loses mass and shrinks it will transition to photoevaporation and will continue to be stripped by photoevaporation until the entire envelope is lost (D). (iii) Finally, a planet born with a core mass of $\sim 8~$M$_\oplus$ will begin life undergoing photoevaporation, and its mass-loss remain photoevaporative throughout (e.g. track E in figure~\ref{fig:XvsM}). Importantly, the general trends mean a planet that is stripped by core-powered mass-loss has likely only ever experienced core-powered mass-loss. Whereas a planet stripped by photoevaporation could have experienced an early phase of core-powered mass loss or could have been entirely photoevaporative. 

We point out two complications to the above summary, demonstrated by trajectories ``A'' and ``B'' in Figure~\ref{fig:XvsM}. Trajectory ``A'' shows a planet which begins with undergoing core-powered mass-loss, then switches to photoevaporation through the penetration limit; however, as it continues to lose mass and as the XUV flux decreases, it enters core-powered mass-loss again due to the heating limit. If it continues to lose mass, it will again transition back to a photoevaporative outflow. Alternatively, trajectory ``B'' shows a planet that begins in core-powered mass-loss before transitioning to photoevaporation at an envelope mass fraction of $\sim 1$\% at the age of a few hundred Myr; however, if it remained with a mass-fraction of $\sim 1$\%, as the XUV flux drops the transition to photoevaporation shifts to smaller envelope mass fractions resulting in the planet switching back to core-powered mass-loss as it evolves over time. If it loses enough mass through core-powered mass-loss, it could again return to a photoevaporative outflow. 

Thus, while possible for a planet to switch multiple times in its life, the standard outcome is a planet either undergoes exclusively core-powered mass-loss (low masses, high temperatures) or photoevaporation (moderate core masses and initial envelope mass fractions) or switches from core-powered mass-loss to photoevaporation (moderate core masses and high initial envelope mass fractions). While our above trajectories are indicative, before full evolutionary calculations are performed, we cannot speculate how far down these trajectories an individual planet may make it in a few billion years. For example, planets may ``stall'' on these trajectories when mass-loss becomes evolutionary unimportant.

\subsection{Mass-loss across the exoplanet population}\label{sec:population}

Using our results, we can sketch out an evolutionary pathway for the close-in, low-mass exoplanets population. Assuming the planets are embedded in their parent protoplanetary disc, in similar orbits before disc dispersal, their radii will fill their Bondi radii ($\sim R_B$), or in some cases, their Hill radii. As disc dispersal begins, the rapid depressurisation of their atmospheres will trigger ``boil-off/spontaneous mass-loss'', a period of rapid mass-loss and shrinking \citep[e.g.][]{Ikoma2012,OwenWu2016,Ginzburg2016,Rogers2023c}. Boil-off/spontaneous mass-loss will then transition to a core-powered mass-loss outflow, as the planets are initially large $\sim 10$~R$_\oplus$ \citep[e.g.][]{Owen2020}. Figure~\ref{fig:XvsM}, indicates a ``typical'' planet (i.e. few percent envelope mass fraction, core mass of 4-5 M$_\oplus$ and equilibrium temperature of 1000~K -- e.g. \citealt{Rogers2021,Rogers2023}) will have transitioned to photoevaporation by the age of 100~Myr, and it is likely photoevaporation will dominate its sub-subsequent mass-loss. Hotter planets can remain undergoing core-powered mass loss to higher planet masses and for longer, whereas photoevaporation will typically dominate for cooler planets. 

Since the trajectory of most planets in Figure~\ref{fig:XvsM} will essentially be vertically downwards, planets can start losing mass in the core-powered mass-loss regime and transition into photoevaporation to be completely stripped or remain undergoing core-powered mass-loss until their envelopes are completely removed at high equilibrium temperature and low core masses. However, as discussed above, it is rare to find a scenario where a photoevaporting planet transitions back to core-powered mass-loss during its mass-loss history and subsequently to be completely stripped. Future evolutionary calculations incorporating both models could explore the possibility that very low-mass, temperate planets can transition between the two mass-loss regimes multiple times during their lifetime. 
\begin{figure}
    \centering
    \includegraphics[width=\columnwidth]{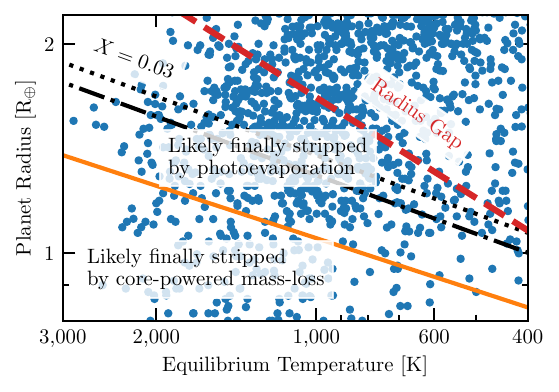}
    \caption{The radii of small detected transiting planets shown as a function of equilibrium temperature. Those planets where the final hydrogen-dominated atmosphere is likely removed by photoevaporation lie above the solid, yellow line and the super-earths where the final stripping likely proceeded by core-powered mass-loss lie below the solid line. The black lines correspond to planets that transitioned from core-powered mass-loss to photoevaporation when their envelope mass fractions equalled 0.03 at 3 Myr (black dotted) and 100 Myr (black dot-dashed). The position of the radius-gap from \citet{Owen2017} is shown as the red dashed line. Exoplanet data was downloaded from the NASA exoplanet archive \citep{NASAarchive} on 22/3/2023.}
    \label{fig:stripped_cores}
\end{figure}

{\rc Inferences about which mechanism removes that majority of the exoplanet populations' mass must await full evolutionary calculations. However, one of the insights we can infer from our work is which mass-loss mechanism is responsible for the {\it final }``stripping'' of the envelopes}. We accomplish this by investigating the core masses at which core-powered mass-loss transitions to photoevaporation for a negligible envelope mass fraction (specifically $10^{-4}$). Unsurprisingly, this transition scales with $R_p/R_B$, and we show this boundary in terms of planet equilibrium temperature and radius compared to the observed exoplanet population in Figure~\ref{fig:stripped_cores}. This indicates that the observed super-earth planet population contains a significant faction of planets where the final removal of the H/He envelope was controlled by either photoevaporation or core-powered mass-loss. However, since the exoplanet radius gap borders the photoevaporative region, it is likely that photoevaporation was responsible for the mass loss of observed super-earths located in its direct vicinity. For sub-Neptunes with hydrogen-dominated atmospheres, planets just above the radius-gap host a few-percent atmospheres by mass \citep[e.g.][]{Wolfgang2015,Owen2017}. Thus, even though core-powered mass-loss and photoevaporation can create the observed radius gap in isolation \citep[e.g][]{Rogers2023b}; the black lines in Figure~\ref{fig:stripped_cores} indicate it was likely that photoevaporation was responsible for the final carving of the radius gap, setting its topography observed today. {\rc Although we emphasise this does not mean photoevaporation removed the majority of the exoplanet population's envelope mass. }
This is because, the planetary cores  straddled by the black lines are expected to have transitioned from core-powered mass-loss to photoevaporation once they reached envelope mass fractions of a few percent on 10-100 Myr timescales.

Distinguishing between the mass-loss mechanisms responsible for the final ``stripping'' of the envelopes is important as photoevaporation and core-powered mass-loss may imprint different final atmospheric compositions during the removal of the last amounts of hydrogen. For example, core-powered mass-loss can leave small residual hydrogen in the atmosphere \citep[e.g.][]{Misener2021}, and photoevaporation can drag heavy elements along with it \citep[e.g.][]{Zahnle1986}. This opens up the possibility for future observational tests of these two mass-loss scenarios.

\subsection{Role of X-rays}\label{sec:Xrays}
\begin{figure}
    \centering
    \includegraphics[width=\columnwidth]{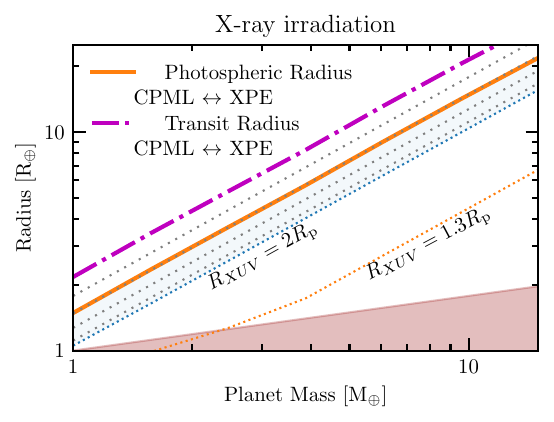}
    \caption{The same as Figure~\ref{fig:Single_model}, but instead of EUV photons penetrating the outflow, X-ray photons with a cross-section of $\sigma_{\rm XUV}=3\times10^{-22}$~cm$^2$ is used. This shifts the transition to photoevaporation to slightly larger planetary radii.}
    \label{fig:Xrays}
\end{figure}
Young stars emit a significant fraction of their XUV output in the X-rays \citep[e.g.][]{Jackson2012,Chadney2015,King2021}. Thus, photoevaporation can be driven by the X-rays rather than the EUV as predominately assumed in our previous calculations. While the controlling physics is similar to the penetration limit for energy-limited flows described above\footnote{There is no ionization recombination equilibrium in the case of soft X-ray irradiation as the X-rays typically remove a K-shell electron rather than valence electrons, and thus X-ray ionization does not increase the transparency of gas to X-ray irradiation.}, the major difference is that the cross-section to the absorption of X-rays is significantly smaller $\sim 10^{-22}$~cm$^2$. This increases the penetration depth of X-rays, resulting in photoevaporation taking over from core-powered mass-loss at larger planetary radii than EUV irradiation. To assess the impact, we remake Figure~\ref{fig:Single_model} for the case of X-ray irradiation in Figure~\ref{fig:Xrays}. This shows the fundamentals are similar, just the transition is shifted to slightly larger radii due to the logarithmic dependence on the absorption cross-section to high-energy photons (Equation~\ref{eqn:penetrate_solution1}).  One major difference is the reduction of parameter space occupied by ``enhanced'' X-ray photoevaporation, as the higher penetration of X-ray photons results in $R_{\rm XUV}$ sitting closer to the planet's radius, a result previously discussed in \citet{Owen2012}. However, as discussed in previous works on X-ray photoevaporation, the isothermal outflow approximation can be fairly poor \citep{Owen2012}. Thus, further simulation work is required to explore the transition between X-ray photoevaporation and core-powered mass loss.

\subsection{Limitations and directions for future work}
Our aim here has been to lay the theoretical foundations to assess the role core-powered mass-loss and photoevaporation play in shaping the exoplanet population.
However, to understand the basics of physics shaping the problem, we have simplified the models. {\rc We consider the bolometrically heated, energy-limited, and recombination-limited regions to be distinct and sharply transition from one another (including sharp transitions in the mean-molecular weight).  In reality, these boundaries are known to be smooth \citep[e.g.][]{Owen2016}, occurring over several scale heights.} {\rcc In particular, in the case where our modelled outflow transitions from a molecular, bolometrically heated outflow, to a fully ionized outflow in recombination-ionization equilibrium, we are neglecting the photo-dissociation region, where molecules are thermally or photo-dissociated. This transition will produce a region where the temperature and mean-molecular weight smoothly changes. While this will change the exact value of $R_p/R_B$ where the transition from core-powered mass-loss to photoevaporation occurs, its logarithmic dependence on planetary and outflow properties mean the basic scalings and inferences we've identified are expected to be robust to our assumptions. }

Furthermore, we treat photoevaporation as either an energy-limited outflow modelled with a constant sound speed or recombination-limited. We also do not attempt to smoothly transition between the two cases nor consider how efficiency evolves with planetary properties. We treat core-powered mass loss as an isothermal outflow occurring at the planet's equilibrium temperature.
This approximation neglects the fact that variations in opacity between the infrared and optical wavelengths can lead to heating of the approximated isothermal region above the nominal equilibrium temperature, which could lead to faster mass-loss rates. {\rc The isothermal assumption also neglects $P{\rm d}V$ cooling in both the bolometrically heated and photoevaporative regions, which can weaken the outflows. Specifically, $P{\rm d}V$ cooling in the bolometrically heated region will lower its scale height, allowing XUV radiation to penetrate closer to the planet's photosphere. This would cause planets to transition from core-powered mass-loss to photoevaporation when they are of lower density than the isothermal calculations imply. This is likely an important correction for low-density planets, whose scale height is already a non-negligible fraction of the planet's radius.    }

Additionally, the treatment of core-powered mass-loss in this paper also neglects the energy limit \citep[e.g.][]{Ginzburg2018}, where the cooling luminosity at the radiative-convective boundary is insufficient to resupply gas to $R_p$. We also do not model the transition from boil-off/spontaneous mass-loss \citep[e.g.][]{Ikoma2012,OwenWu2016,Ginzburg2016} to core-powered mass-loss and photoevaporation explicitly. Realistic radiation-hydrodynamic simulations \citep[e.g.][]{GarciaMunoz2007,MurrayClay2009,Owen2012,Kubyshkina2018a,Kubyshkina2018b}, indicate while the above represents a broad-brush approach to the problem, it neglects many of the details which can change the mass-loss rates by order unity factors. In particular, it's worth reiterating that constant efficiency energy-limited photoevaporation is inconsistent with the slope of the radius-gap, while the radiation-hydrodynamic simulations are consistent \citep{VanEylen2018}. 

Furthermore, while we have used our results to sketch various evolutionary histories, identifying the boil-off and core-powered mass-loss dominates early, before switching to photoevaporation in many cases, without evolutionary calculations, it's unclear where different amounts of mass-loss occur. Nonetheless, since pure photoevaporation \citep[e.g.][]{Owen2017,Jin2018,Wu2019,OwenAdams2019,Rogers2021,Rogers2023b}  and core-powered mass-loss models \citep[e.g.][]{Ginzburg2018,Gupta2019,Gupta2020,Gupta2022} give conceptually similar results for the origin and physical properties of the close-in exoplanet populations, we expect these results to be robust. 

Finally, we've identified that observed terrestrial planets can either have been getting their final atmospheric stripping by core-powered mass-loss or photoevaporation. Properties of the atmospheres of these terrestrial planets are beginning to be observed; for example, LHS 3844b \citep{Kreidberg2019}, GJ 1252b \citep{Crossfield2022} and Trappist-1b \citep{Greene2023} all sit in the likely to have been finally striped by photoevaporation. More theoretical work on the residual atmospheres left behind, in concert with continued observations, should be able to test the roles of mass loss from hydrogen-dominated primary atmospheres in controlling the secondary atmospheres of hot rocky exoplanets. 

\section{Conclusions}

We have studied how atmospheric escape from close-in planets transitions from core-powered mass-loss to photoevaporation. By focusing on (semi-)analytic methods, we have provided physical insights that should help guide future, expensive radiation-hydrodynamic simulations necessary to fully map out the mass-loss rates across the planet and stellar parameters. Our main results are as follows:
\begin{enumerate}
    \item A planetary outflow will occur in the core-powered mass-loss regime if it cannot be penetrated by XUV photons interior to the Bondi radius or if the XUV photons provide insufficient heating. Across most of the planetary parameter space, the penetration of XUV photons interior to the Bondi radius sets the transition between the two outflow regimes. 
    \item The transition between core-powered mass-loss occurs at roughly constant $R_p/R_B$ (or escape parameter), where $R_B$ is the sonic point of a core-powered mass-loss outflow. This ratio takes a value of roughly $1/5-1/9$, with the exact value only being logarithmically sensitive to stellar and planetary parameters.
    \item Thus, core-powered mass-loss dominates for hot, puffy planets, while photoevaporation dominates for denser cooler planets. Where typically core-powered mass-loss transitions to recombination limited EUV photoevaporation.
    \item Under most situations, a planet can transition from core-powered mass-loss to photoevaporation as it evolves, but not vice-versa. Meaning a planet that is completely stripped by core-powered mass-loss will only have ever experienced core-powered mass-loss. 
    \item Observed close-in exoplanets cover planets that only ever experienced core-powered mass-loss or photoevaporation or transitioned from core-powered mass-loss to photoevaporation.
    \item Even when the mass-loss is photoevaporative, core-powered mass loss can ``enhance'' photo-evaporation over a significant region of parameter space. 
    \item Observed, rocky terrestrial planets are likely to have been stripped by core-powered mass-loss at high equilibrium temperatures and low mass, whereas they were {\it finally} stripped by photoevaporation at cooler temperatures and higher masses.
    \item Applying our results to the observed super-Earth population indicates that it contains significant fractions of planets where the final removal of the H/He envelope occurred in both regimes.
    \item Photoevaporation was likely responsible for the final carving of the exoplanet radius-valley, setting its topography; however, core-powered mass-loss/boil-off should have played a role earlier in these planet's evolution depending on their initial hydrogen inventories.
\end{enumerate}

Since core-powered mass-loss and photoevaporation both operate in the observed parameter space, work needs to be done to incorporate a combined model into evolutionary calculations to explore exoplanet demographics, similar to the core-powered mass-loss only \citep[e.g.][]{Gupta2019,Gupta2020} and photoevaporation only works \citep[e.g.][]{Wu2019,Rogers2021,Rogers2021b}.

\section*{Acknowledgements}
We are grateful for comments from the anonymous referee, which improved the manuscript. We thank Morgan MacLeod for helpful discussions. This research has been supported by NASA's Exoplanet Research Program (XRP) under grant number 80NSSC21K0392. JEO is also supported by a Royal Society University Research Fellowship.  JEO. has also received funding from the European Research Council (ERC) under the European Union’s Horizon 2020 research and innovation programme (Grant agreement No. 853022, PEVAP). For the purpose of open access, the authors have applied a Creative Commons Attribution (CC-BY) licence to any Author Accepted Manuscript version arising.

\section*{Data Availability}
The work underlying this article will be shared on reasonable request to the corresponding author.



\bibliographystyle{mnras}
\bibliography{example} 



\onecolumn
\appendix

\section{Optical depth into an isothermal density structure}\label{app:opt_depth_calc}
In this appendix, we outline the solution to Equation~\ref{eqn:tau_1_energy}. Adopting the hydrostatic density profile Equation~\ref{eqn:iso_hydrostatic}, Equation~\ref{eqn:tau_1_energy} can be written as:
\begin{equation}
    \int_{R_{\rm XUV}/R_p}^\infty\exp\left[-\frac{2R_B}{R_p}\left(1-\frac{1}{x}\right)\right]{\rm d}x = \frac{\mu_{\rm eq}}{\rho_{\rm phot}\sigma_{XUV}R_p}
\end{equation}
where $x=r/R_p$. The integral on the LHS can be approximately computed using asymptotic integration (specifically Laplace's method), which to first order yields:
\begin{equation}
    \frac{1}{2}\left(\frac{R_p}{R_B}\right)\left(\frac{R_{\rm XUV}}{R_p}\right)^2\exp\left[\frac{2R_B}{R_p}\left(\frac{R_p}{R_{XUV}}-1\right)\right] = \frac{\mu_{\rm eq}}{\rho_{\rm phot}\sigma_{XUV}R_p}
\end{equation}
This approximation is physically equivalent to saying the optical depth is dominated by the last few scale heights, close to $R_{\rm XUV}$. Now using Equation~\ref{eqn:phot_density} we find:
\begin{equation}
    \left(\frac{R_{\rm XUV}}{R_p}\right)^2\exp\left[\frac{2R_B}{R_p}\left(\frac{R_p}{R_{XUV}}-1\right)\right] = \frac{\sigma_{IR}}{\sigma_{\rm XUV}}
\end{equation}
The above transcendental function can be expressed in terms of the Lambert W function, which is multi-valued in the range of interest, indicating the presence of multiple solutions. Thus, writing $R_{\rm XUV}=R_p(1+A)$ and expanding for two solutions, one when $A\ll1$ and $A\gg1$ yields our two desired solutions. For $A\ll1$:
\begin{equation}
    A \approx \frac{R_p}{R_B}\log\left(\sqrt{\frac{\sigma_{XUV}}{\sigma_{IR}}}\right)
\end{equation}
and for $A\gg1$:
\begin{equation}
    A \approx \sqrt{\frac{\sigma_{XUV}}{\sigma_{IR}}} \exp\left(-\frac{R_B}{R_p}\right)
\end{equation}
Given the cross-sections are essentially constants, which solution is appropriate depends on the value of $R_p/R_B$, with the transition at roughly a value of $\sim 10$.

\section{Generalised Isothermal Wind Solutions} \label{app:general_parker}

The classical Parker wind solution only applies in the case of vacuum boundary conditions and the flow remains isothermal all the way to the sonic point. However, if one connects to a photoevaporative flow the isothermal region must drive a higher mass-loss rate than the Parker solution.
The first integral to the isothermal wind problem is given by:
\begin{equation}
    U^2 -\log\left(U^2\right) = \log\left[\left(\frac{r}{R_s}\right)^4\right] + 4\frac{R_s}{r} +c
\end{equation}
The Parker wind solution has $c=-3$; however, solutions with $c<-3$ have higher velocities at a given radius and are the flows that connect onto the photoevaporative solution. Thus the generalised solution is given by:
\begin{equation}
    U = \sqrt{-W\left[-\left(\frac{Rs}{r}\right)^4\exp\left(-c - 4\frac{R_s}{r}\right)\right]}
\end{equation}
where $W$ is the Lambert W function (the -1 branch for super-sonic flows). Therefore, to find the correct outflow profile in the case the transonic solution indicates unphysical super-sonic launching, we solve for the constant $c$, such that the flow velocity is equal to the sound speed at $R_{\rm XUV}$. 

\bsp	
\label{lastpage}
\end{document}